\documentstyle[epsf,12pt]{article}
\makeatletter
\@addtoreset{equation}{section}
\makeatother
\newcommand{\etab}{\mbox{$\bar{\eta}$}}
\newcommand{\psib}{\mbox{$\bar{\psi}$}}
\newcommand{\xib}{\mbox{$\bar{\xi}$}}
\newcommand{\dsl}{\mbox{$\partial\!\!\! /$}}
\newcommand{\psla}{\mbox{$p\!\!\! /$}}
\newcommand{\qsla}{\mbox{$q\!\!\! /$}}
\newcommand{\Gb}{\mbox{$\bar{G}$}}
\newcommand{\fex}{\mbox{$\exp [(2\psla_{1}+2\psla_{3}+4m)T]$}}
\date{August 26, 1995}
\begin{document}
\title{Stochastic Coupling of Fermions}
\author{{\sc Jan \v{R}\'{\i}dk\'{y}}        \\
{\it Institute of Physics, Academy of Sciences of the Czech Republic}
\thanks{Postal address: Na Slovance 2, 180~40 Prague 8, Czech Republic;
E-mail address: ridky@fzu.cz}}
\begin{titlepage}
\maketitle
\begin{abstract}
The stochastic quantization of the fermion field is performed
starting from Dirac equations. The statistical properties of
stochastic terms in Langevin equations are described by
explicit formulae of a Markov process. The interaction of the
field is introduced as correlation of the stochastic terms.
In the long time limit free fermions disappear and proper
combinations of field components propagate as a scalar boson
field. The existence and uniqueness of the long time limit
is proved in the first order approximation of stochastic
Liouville equation.
\end{abstract}
\end{titlepage}

%\newpage
\section{Introduction}
\label{secI}

     The method of stochastic quantization \cite{ParS,ParW} starts from
Euclidean field equations amended by addition of a random stochastic
variable. Thus the change of the field during an infinitesimal increment
of stochastic time {\em t} which is considered to be an auxiliary
variable is described by field equation and stochastic term. The massive
scalar field is described e.g. by
\begin{equation}
{ \partial \over \partial t}  \phi (x,t) =
(- \partial^2 + m^2) \phi (x,t)\, + \, \xi (x,t)
\label{I.1}
\end{equation}
where $x$ are coordinates in Euclidean space. (\ref{I.1}) can be also
regarded as Langevin equation with drift term equal to Klein-Gordon
operator. The statistical properties of the fluctuation term $\xi(x,t)$
are given. One particular solution of (\ref{I.1}) is of no special
interest but the mean value of the correlation
$\langle\phi(x,t),\phi(x^{'},t)\rangle$
averaged over all possible fluctuations is equal in the long time limit
(i.e. $t \rightarrow \infty$ ) to the usual result of path integral
quantization of the Euclidean field theory under consideration. One has
of course to prove that the long time limit exists and it is unique.

     In the standard approach the fluctuation term corresponds to white
noise - i.e. it is a random Gaussian variable with autocorrelation
equal to \begin{equation}
\langle\xi(x,t)\xi(x',t')\rangle\; = \; c \, \delta (x-x')\, \delta (t-t')
\label{I.2}
\end{equation}
The finite width of the Gaussian distribution of $\xi$ in $t$ variable
in (\ref{I.1}) leads to Pauli-Villars regularization of propagators
\cite{JAP3}.

     Stochastic quantization of boson fields is the easier one.
Stochastic quantization of fermion fields is more elaborate and there
are two approaches. The method proposed by Breit, Gupta and Zaks
\cite{BGZ} starts also from Klein-Gordon operator as in (\ref{I.1}) and
the fermion propagator results from special statistical properties of
the fluctuation term. In this approach
\[{ \partial \over \partial t}  \psi (x,t) =
(- \partial^2 + m^2) \psi (x,t) + \eta (x,t) , \]
\[{ \partial \over \partial t}  \psib\ (x,t) =
\psib\ (x,t)(- \partial^2 + m^2) + \etab\ (x,t) \]
and
\begin{equation}
\langle\eta(x,t)\etab\ (x',t')\rangle \; = \; c \, ( -i \dsl\ + m )\,
\delta (x-x') \, \delta (t-t').
\label{I.3}
\end{equation}
Thus quantization of boson and fermion fields is put on the same footing
and the stochastic time has in both two cases the same dimension
$[t]=m^{-2}$. This is convenient for perturbative description of
interacting fermion fields.

     The second method proposed by Fukai et al. \cite{JAP5} starts from
Dirac operators
\[{ \partial \over \partial t}  \psi (x,t) =
( i \dsl\ - m ) \psi (x,t) + \xi (x,t) , \]
\begin{equation}
\label{I.4}
\end{equation}
\[{ \partial \over \partial t}  \psib\ (x,t) =
\psib\ (x,t)( -i \dsl\ - m ) + \xib\ (x,t) \]
and the stochastic time {\em t} has the real time dimension. It seems to
be quite difficult to implement this method in the case of interacting
fermion fields within the framework of standard perturbation approach.
However, it allows to quantize fermions in such a way that correlations
of proper fermion field combinations are equal to autocorrelations of a
boson field. Such procedure is the subject of this paper.

     The idea is to introduce the interaction between fermion fields via
correlation of fluctuation terms. Direct consequence of simple correlations
of the type (\ref{I.2}) is locality of resulting theory. As the correlations
of the fluctuation terms are related to those of the field components, more
elaborate probability distribution of stochastic terms can lead, in addition
to ensuring the locality, also to dynamical consequences.
To achieve this it is necessary to use one multivariate
distribution governing behavior of all fluctuation terms in Langevin
equations at given stochastic time {\em t}. The sequence of these
distributions will form a Markov process. It is evident that
random variables in (\ref{I.4}) have to be of anticommuting nature.
Markov processes of Grassmann numbers are constructed and studied in the
Section~\ref{secM} of the paper. Lowest order propagator based on
Langevin equations with correlated fluctuation terms is calculated in
the Section~\ref{secF}. The existence and the uniqueness of the long
time limit of the studied system is proved in first order approximation
in the Section~\ref{secL}. The Section~\ref{secC} of the paper is devoted to
conclusions and discussion.

\section{Markov chains of Grassmann numbers}
\label{secM}

     Classical Markov processes \footnote{An excellent introduction into
Markov processes and stochastic differential equations can be found in
\cite{Kemp}.} are based on the notion of probability distribution.
However, there is no such concept as probability measure in Grassmann
number calculus. Hence one has to adopt a more general approach outlined e.g.
in \cite{Berez} and also used in \cite{JAP5}. It is possible to use as a
probability distribution of Grassmann variables $\xi, \xib\ $
(henceforth it will be dealt always with pairs of Grassmann numbers) a
function which can be generally written as
\begin{equation}
{P} ( \xi, \xib\ ) =
c_{0} + c_{1} \xi + c_{2} \xib\ + c_{3} \xi \xib\
\label{M.1}
\end{equation}
and such that the integral
\begin{equation}
\int f(\xi,\xib\ ) {P} ( \xi, \xib\ ) \; d \xi d \xib\ \equiv
\;\; \langle f(\xi,\xib\ ) \rangle
\label{M.2}
\end{equation}
gives correct expectation values $\langle f(\xi,\xib)\rangle$ for any
function $f$. It is reasonable to assume that ${P}$ commutes with any
Grassmann number. The conventions of Grassmann variables integration and
derivations (left-derivatives) are the same as those used in \cite{JAP5},
\begin{equation}
\int d \xi = 0 \, ,\; \int \xi \; d \xi = i \, , \;
\int d \xib\ = 0 \, , \; \int \xib\ d \xib\ = i
\label{M.3}
\end{equation}
so that e.g.
\[ \int \xi \xib\ d \xi d \xib\ = 1. \]
\begin{equation}
{ \partial \over \partial \xi} ( \xi \xib\ )  = \xib\ \; , \;\;
{ \partial \over \partial \xib\ } ( \xi \xib\ )  =
 - { \partial \over \partial \xib\ } ( \xib\ \xi )  = - \xib\ .
\label{M.4}
\end{equation}
Classical Markov process is given by the probability distribution
${P}( y_{1},t_{1} )$ of stochastic variable $y_{1}$ at certain time
$t_{1}$ ( with the abbreviation $y_{1} \equiv y(t_{1})$ ) and the
transition probability ${T}(y_{2},t_{2} \mid y_{1},t_{1} )$
which determines the probability distribution of $y$ at any time $t_{2}
\geq t_{1}$
\begin{equation}
{P}( y_{2},t_{2} ) = \int {T}(y_{2},t_{2} \mid y_{1},t_{1} )
{P}( y_{1},t_{1} ) \, d y_{1}
\label{M.5}
\end{equation}

The transition probability satisfies Chapman-Kolmogorov equation for any
$t_{3} > t_{2} > t_{1}$
\begin{equation}
{T}(y_{3},t_{3} \mid y_{1},t_{1}) = \int
{T}(y_{3},t_{3} \mid y_{2},t_{2}) {T}(y_{2},t_{2} \mid
 y_{1},t_{1})
\, d y_{2}
\label{M.6}
\end{equation}
and it is normalized
\begin{equation}
\int {T}(y_{2},t_{2} \mid y_{1},t_{1}) \, d y_{2} = 1.
\label{M.7}
\end{equation}
One of the most simple stochastic processes one can use is the
Orenstein-Uhlenbeck process. The natural step therefore will be to construct
Grassmann variable analog to this process which is the unique stationary
Gaussian Markov process and it is defined by
\begin{equation}
 {P} ( y_{1} ) = \frac{1}{\sqrt{2 \pi \sigma_{\scriptscriptstyle E}}} \;\;
e^{-\frac{y_{1}^{2}}{2 \sigma_{\scriptscriptstyle E}}},
\label{M.8a}
\end{equation}
\begin{equation}
 {T}(y_{2},t_{2} \mid y_{1},t_{1})  =
\frac{1}{\sqrt{2 \pi \sigma_{\scriptscriptstyle E}(1-e^{-2M \tau})}} \;\;
\exp \left[ -\frac{(y_{2}-y_{1}e^{-M\tau})^{2}}
{2 \sigma_{\scriptscriptstyle E}(1-e^{-2M \tau})} \right] \, ,
\label{M.8b}
\end{equation}
where $\tau = t_{2} - t_{1}$.

     The multivariate distributions are needed but for simplicity in the
following the functions of only one pair of Grassmann variables
$\xi$ and $\xib\ $ will be used. All necessary mathematical properties
can be demonstrated by means of such functions. One can take as Grassmann
Gaussian probability distribution
\begin{equation}
{P}( \xi , \xib\ ) = - 1 - \xib\ \xi
\label{M.9}
\end{equation}
and readily check that the normalization condition holds
\begin{equation}
\int {P}( \xi , \xib\ ) \; d \xi d \xib\ =  1.
\label{M.10}
\end{equation}
If one tries as transition probability distribution
\begin{equation}
{T}( \xi_{2} , \xib_{2} \mid \xi_{1} , \xib_{1} ) =
- 1 - \xib_{2}\xi_{2} - e^{\pm M(t_{2}-t_{1})}
[ \xi_{2} \xib_{1} - \xib_{2} \xi_{1} ]
\label{M.11}
\end{equation}
then the following equations hold
\begin{equation}
\int {T}( \xi_{2} , \xib_{2} \mid \xi_{1} , \xib_{1} )
{P}(\xi_{1},\xib_{1}) \;
d \xi_{1} d \xib_{1} = {P}(\xi_{2},\xib_{2}),
\label{M.12}
\end{equation}
\begin{equation}
\int {T}( \xi_{3} , \xib_{3} \mid \xi_{2} , \xib_{2} )
{T}( \xi_{2} , \xib_{2} \mid \xi_{1} , \xib_{1} ) \;
d \xi_{2} d \xib_{2} = {T}( \xi_{3} , \xib_{3} \mid \xi_{1} ,
\xib_{1}) ,
\label{M.13}
\end{equation}
provided that $t_{3} > t_{2} > t_{1}$ , and finally
\begin{equation}
\int {T}( \xi_{2} , \xib_{2} \mid \xi_{1} , \xib_{1})
\; d \xi_{2} d \xib_{2} = 1 .
\label{M.14}
\end{equation}
They correspond to (\ref{M.5},\ref{M.6}) and (\ref{M.7}) respectively.
However, there is one significant difference - the exponent in
(\ref{M.11}) can be also positive in contrast to (\ref{M.8b}) where the
positive exponent in $t$ dependence would lead to complex ${T}$
values.

     The analogy of (\ref{M.9},\ref{M.11}) with the classical process
can be extended even further. The equation
\begin{equation}
\frac{\partial {T}}{\partial \tau} \; = \; \pm M \left[
\frac{\partial}{\partial \xi_{2}} (\xi_{2}\,{T})\:+\:
\frac{\partial}{\partial \xib_{2}} (\xib_{2}\,{T})\:+2\:
\frac{\partial^{2}\:{T}}{\partial \xib_{2}\:\partial \xi_{2}}
\right] \; ,
\label{M.15}
\end{equation}
is the forward Kolmogorov equation
for Grassmann variable Markov process (\ref{M.11}), while the forward
Kolmogorov equation of the classical process
(\ref{M.8a},\ref{M.8b}) reads
\[
\frac{\partial {T}}{\partial \tau} \; = \; - M \left[
\frac{\partial}{\partial y_{2}} (y_{2}\,{T})\:+\:
\frac{\partial^{2}\:{T}}{\partial y_{2}^{2}} \right] \; ,
\]
The backward Kolmogorov equation for (\ref{M.11}) has no analogy in
corresponding classical equation. The full analogy is restored for
\begin{equation}
{T^{'}}(\xi_{2},\xib_{2} \mid \xi_{1},\xib_{1})\:=\:
(e^{\pm 2M\tau}\,-\,1)\:-\:(\xib_{2}\,-\,e^{\pm M\tau}\,\xib_{1})
(\xi_{2}\,-\,e^{\pm M\tau}\,\xi_{1}).
\label{M.16}
\end{equation}
This transition probability distribution is the full analog to that of
classical process, including both the forward and backward Kolmogorov
equation. In contrast with classical Orenstein-Uhlenbeck process, which
is the unique stationary Gaussian Markov process, here one has more
stationary processes which can be called Gaussian. This is only natural as
one deals with truncated Taylor series. The Kolmogorov equations of
the process of anticommuting variables (\ref{M.16}) have the same
asymptotics for $\tau \rightarrow \pm \infty $ as the classical process.
In the following the analog of simpler (\ref{M.11}) will be used.

The stochastic variables which will enter the Langevin
equations will be represented by quartets of
functions which ascribe to each point in Euclidean space-time $x$ and to
each value of stochastic time $t$ four random Grassmann numbers. To achieve
the desired result - coupling of four fermions - it turns out that
the probability distributions of these variables have to be functions
of two pairs of such stochastic
variables ($\xi_{\mu}(x,t) \, ,\,\eta_{\mu}(x,t)$ and
$\xib_{\mu}(x',t) \, ,\,\etab_{\mu}(x',t)$, where $\mu = 0, 1, 2, 3$).
These pairs will be
defined on a product of two Euclidean spaces $x \times x'$ and their
distribution in this space will be described by $\delta$-function
$\delta(x=x')$. Further one has to suppose that these "quartets of quartets"
have nonzero product of all their components - i.e. all 16 Grassmann
numbers are different. Thus the probability distribution in $x'$
becomes for any $x$
\begin{equation}
{\cal P}(\xi (x',t),\xib (x',t),\eta (x',t),\etab (x',t))\mid_{x}  \equiv
{P}(\xi (x,t),\xib (x,t),\eta (x,t),\etab (x,t))\,
\delta (x - x')
\label{M.17}
\end{equation}
and the transition probability distribution as a function of $y'$ will be
\begin{eqnarray}
\label{M.18}
\lefteqn{{\cal T}
(\xi (y',t_{2}),\xib (y',t_{2}),\eta (y',t_{2}),\etab (y',t_{2})
\! \mid \!
\xi (x,t_{1}),\xib (x,t_{1}),\eta (x,t_{1}),\etab (x,t_{1}))\mid_{y}
\: \equiv }\\
&& {T}
(\xi (y,t_{2}),\xib (y,t_{2}), \eta (y,t_{2}),\etab (y,t_{2})
\! \mid \!
\xi (x,t_{1}),\xib (x,t_{1}),\eta (x,t_{1}),\etab (x,t_{1}))
\,\delta (y - y')\,     \nonumber
\end{eqnarray}
for any $x, y$.
Both in (\ref{M.17}) and (\ref{M.18}) the greek indices are omitted. In
rigorous approach the probability distributions in Euclidean space in
(\ref{M.17}) and (\ref{M.18}) should be e.g. of the type listed in
(\ref{M.8a}). One
can consider the $\delta$-functions in (\ref{M.17},\ref{M.18}) to
result from the limit $\sigma_{\scriptscriptstyle E} \rightarrow 0$
in (\ref{M.8a}). Simultaneously the $\sigma_{\scriptscriptstyle E}$
parameter disappears. In fact one allows in (\ref{M.1}) as
$c$-coefficients not only $c$-functions but also $c$-distributions.

     As the random Grassmann variable distribution
${P}(\xi (x\!,\!t),\xib (x\!,\!t),\eta (x\!,\!t),\etab (x\!,\!t))$ in
(\ref{M.17}) at given stochastic time $t_{1}$ the following function
will be taken as ansatz
\begin{equation}
{P}(\xi^{1},\xib^{\,1},\eta^{1},\etab^{\,1})\:  = \:
\prod_{\mu =0}^{3}\, \eta_{\mu}^{1}
\prod_{\rho =0}^{3}\, \xib_{\rho}^{\,1}
\prod_{\sigma =0}^{3}\, \xi_{\sigma}^{1}
\prod_{\nu =0}^{3}\, \etab_{\nu}^{\,1} \, + \, \sum_{i,j=0}^{3} \, (
\prod_{\stackrel{\mu =0}{\mu \neq i}}^{3}\, \eta_{\mu}^{1}
\prod_{\stackrel{\rho =0}{\rho \neq j}}^{3}\, \xib_{\rho}^{\,1}
\prod_{\stackrel{\sigma =0}{\sigma \neq j}}^{3}\, \xi_{\sigma}^{1}
\prod_{\stackrel{\nu =0}{\nu \neq i}}^{3}\, \etab_{\nu}^{\,1})
\label{M.19}
\end{equation}
where the greek indices of $\xi ,\xib ,\eta$ and $\etab $ on the
left hand side of (\ref{M.19}) are omitted and the $x$ dependence
is omitted on both sides. The
stochastic time dependence is indicated by upper right
index, e.g. $\eta(t_{1}) \equiv \eta^{1}$.
Transition probability distribution on right hand side of (\ref{M.18})
written in the same notation reads
\begin{eqnarray}
\lefteqn{{T}(\xi^{2},\xib^{\,2},\eta^{2},\etab^{\,2} \mid
\xi^{1},\xib^{\,1},\eta^{1},\etab^{\,1}) \: = \:
{P}(\xi^{2},\xib^{\,2},\eta^{2},\etab^{\,2}) \: + }  \nonumber \\
& & \mbox{} + \, k \, \sum_{i,j=0}^{3} \: (-1)^{i+j}\: \:
\prod_{\stackrel{\mu =0}{\mu \neq i}}^{3}\, \eta_{\mu}^{2} \:
\prod_{\stackrel{\rho =0}{\rho \neq j}}^{3}\, \xib_{\rho}^{\,2} \:
\prod_{\sigma =0}^{3}\, \xi_{\sigma}^{2} \:
\prod_{\nu =0}^{3}\, \etab_{\nu}^{\,2} \: \cdot
\,\xib^{\,1}_{j} \, \eta^{1}_{i} \: +                       \\
& & \mbox{} + \, k \, \sum_{i,j=0}^{3} \: (-1)^{i+j}\: \:
\prod_{\mu =0}^{3}\, \eta_{\mu}^{2} \:
\prod_{\rho =0}^{3}\, \xib_{\rho}^{\,2} \:
\prod_{\stackrel{\sigma =0}{\sigma \neq j}}^{3}\, \xi_{\sigma}^{2} \:
\prod_{\stackrel{\nu =0}{\nu \neq i}}^{3}\, \etab_{\nu}^{\,2} \: \cdot
\, \etab^{1}_{i}  \, \xi^{\,1}_{j} \: ,     \nonumber
\label{M.20}
\end{eqnarray}
where
\[ k\, =\, e^{\pm M\tau} \: ,\: \tau \, = \, t_{2} \, - \, t_{1} \: ,
\; t_{2} \, > \, t_{1} \, . \]
     Distributions (\ref{M.17}) and (\ref{M.18}) obey the relations
(\ref{M.12}) - (\ref{M.14}) amended by corresponding space-time
integrations preceeding the Grassmann variable integrations. This
stochastic process leads to correlations
\begin{equation}
\langle \xib_{\rho}(x_{1},t_{1})\, \eta_{\mu} ({x'}_{1},t_{1})
\, \etab_{\nu}(x_{2},t_{2})\, \xi_{\sigma}({x'}_{2},t_{2}) \rangle \, =
c_{\scriptscriptstyle E}\:\delta (x_{1} - {x'}_{1})\:\delta (x_{2} - {x'}_{2})
\: \delta_{\mu \, \nu} \: \delta_{\rho \, \sigma} \:
e^{\pm M(t_{2}-t_{1})}
\label{M.21}
\end{equation}
where $t_{1} \leq t_{2}$. The constant
$c_{\scriptscriptstyle E}\, =\, 1,\:[c_{\scriptscriptstyle E}]\,=\,m^{-8}$
is introduced to adjust
dimensions. As the stochastic variables will disappear after integrations
in all applications in Sections 3 and 4 it is simpler to consider them to be
dimensionless and to adjust dimensions in Langevin equations by another
constant $r_{\scriptscriptstyle D}$. This constant can be also used to
readjust the numerical value of $c_{\scriptscriptstyle E}$ so that it
remains equal to $1$ even after change of units.

All expectation values of single Grassmann variables or incomplete pairs
$\eta \, \xib$ or $\etab \, \xi$ are are equal to zero, e.g.
\[ \langle\, \xib_{\mu}(x_{1},t_{1})\, \rangle\: =\:
\langle\, \eta_{\mu} (x_{1},t_{1})\, \rangle\: = \:
\langle\, \etab_{\mu}(x_{1},t_{1})\, \eta_{\nu}(x_{1},t_{1})\, \rangle\:
=\mbox{} \]
\begin{equation}
\label{M.22}
\end{equation}
\[ \mbox{} =\: \langle\, \xib_{\mu}(x_{1},t_{1})\, \eta_{\nu} (x_{1},t_{1})\,
 \etab_{\rho}(x_{1},t_{1})\, \rangle\: =\: etc. \: = \: 0. \]
Also if there is not complete pair $\eta \, \xib$ or $\etab \, \xi$
at any time $t$ the correlation is equal to zero, e.g.
\begin{equation}
\langle\, \xib_{\rho}(x_{1},t_{1})\, \eta_{\mu} (x_{2},t_{2})
\, \etab_{\nu}(x_{3},t_{3})\, \xi_{\sigma}(x_{3},t_{3})\, \rangle \: = \: 0
\label{M.23}
\end{equation}
for $t_{1}\, < \, t_{2}\, \leq \, t_{3}$. The expectation values are zero
also in case of the combinations like
\[ \langle\, \xib_{\rho}(x_{1},t_{1})\, \eta_{\mu} (x_{1},t_{1})
\, \xib_{\sigma}(x_{2},t_{2})\,\eta_{\nu}(x_{2},t_{2})\, \rangle\: =\mbox{} \]
\begin{equation}
\label{M.24}
\end{equation}
\[ \mbox{}=\: \langle\, \xi_{\rho}(x_{1},t_{1})\, \etab_{\mu}(x_{1},t_{1})
\, \xi_{\sigma}(x_{2},t_{2})\,\etab_{\nu}(x_{2},t_{2})\, \rangle\: =\: 0 \]
assuming $t_{1}\, \leq \, t_{2}$. Also the mean value
\begin{equation}
\langle\, \xib_{\rho}(x_{1},t_{1})\, \xi_{\mu} (x_{2},t_{2})
\, \etab_{\nu}(x_{3},t_{3})\, \eta_{\sigma}(x_{4},t_{4})\, \rangle \: = \: 0
\label{M.25}
\end{equation}
whenever the times $t_{1},\, t_{2},\, t_{3}$ and $t_{4}$ are not equal
each other.

     The higher moments are reduced to the product of the lower ones,
e.g.
\begin{eqnarray}
\langle
\xib_{\rho}(x_{1},t_{1})\eta_{\mu} (x_{1},t_{1})
\etab_{\nu}(x_{2},t_{2})\xi_{\sigma}(x_{2},t_{2})
\xib_{\alpha}(x_{3},t_{3})\eta_{\gamma} (x_{3},t_{3})
\etab_{\delta}(x_{4},t_{4})\xi_{\beta}(x_{4},t_{4}) \rangle = \mbox{}
 \nonumber \\
\langle
\etab_{\rho}(x_{1},t_{1})\xi_{\mu} (x_{1},t_{1})
\xib_{\nu}(x_{2},t_{2})\eta_{\sigma}(x_{2},t_{2})
\xib_{\alpha}(x_{3},t_{3})\eta_{\gamma} (x_{3},t_{3})
\etab_{\delta}(x_{4},t_{4})\xi_{\beta}(x_{4},t_{4}) \rangle
 \\
=
\langle \etab_{\rho}(x_{1},t_{1})\xi_{\mu} (x_{1},t_{1})
\xib_{\nu}(x_{2},t_{2})\eta_{\sigma}(x_{2},t_{2}) \rangle
\langle \xib_{\alpha}(x_{3},t_{3})\eta_{\gamma} (x_{3},t_{3})
\etab_{\delta}(x_{4},t_{4})\xi_{\beta}(x_{4},t_{4}) \rangle \nonumber
\label{M.26}
\end{eqnarray}
provided that $t_{1}\,<\,t_{2}\,<\,t_{3}\,<\,t_{4}\,$.

\section{Coupled fermions}
\label{secF}

     The subject of the interest will be a fermion field $\psi$ with no
other interaction except the one represented by Markov process
(\ref{M.17}),(\ref{M.18}). The Langevin equations in this case are
\[{ \partial \over \partial t}  \psi_{\alpha}\, = \,
(i\dsl\ - m)_{\alpha \beta}\, \psi_{\beta}\,+\,
r_{\scriptscriptstyle D}\,(\xi_{\alpha}+\eta_{\alpha}) \]
\begin{equation}
\label{F.1}
\end{equation}
\[{ \partial \over \partial t}  \psib_{\alpha}\, = \,
-\psib_{\beta}\, (i\dsl\ + m)_{\beta\alpha}\,+\,
r_{\scriptscriptstyle D}\,(\xib_{\alpha}+\etab_{\alpha}) \]
where
\begin{equation}
\dsl\ \:=\:\gamma^{\mu}\,\partial_{\mu}\:=\:\gamma^{\mu}\,
{ \partial \over \partial x^{\mu}}
\label{F.2}
\end{equation}
and $\gamma^{\mu}$ are Dirac matrices in Euclidean space, satisfying
\begin{equation}
\{\,\gamma^{\mu}\, ,\,\gamma^{\nu}\, \}\:=\:-\,2\,\delta_{\mu \nu}\,.
\label{F.3}
\end{equation}
The $r_{\scriptscriptstyle D}$ is a constant with dimension $[m^{5/2}]$.
The stochastic variables are dimensionless as discussed in Sec. 2.

     To solve (\ref{F.1}) it is useful to introduce retarded Green functions
\[ G(x,t)\:=\:\Theta (t)\,\int\,\frac{d^{4}p}{(2\pi)^{4}}\,
\exp [-(\psla\ +m)t+ipx]  \]
\begin{equation}
\label{F.4}
\end{equation}
\[ \Gb\ (x,t)\:=\:\Theta (t)\,\int\,\frac{d^{4}p}{(2\pi)^{4}}\,
\exp [-(\psla\ +m)t-ipx]  \]
which satisfy
\[ G(x,t)\:=\:\Gb\ (x,t)\:=\:0\:\:\:\mbox{for}\:\:\:t\,<\,0\, , \]
and
\[{ \partial \over \partial t}G_{\alpha \gamma}(x,t)\, - \,
(i\dsl\ - m)_{\alpha \beta}G_{\beta \gamma}(x,t)\:=\:
\delta_{\alpha \gamma}\,\delta^{4}(x)\,\delta(t) \]
\begin{equation}
\label{F.5}
\end{equation}
\[{ \partial \over \partial t}\Gb_{\alpha \gamma}(x,t)\, - \,
\Gb_{\alpha \beta}(x,t)(-i\dsl\ - m)_{\beta \gamma}\:=\:
\delta_{\alpha \gamma}\,\delta^{4}(x)\,\delta(t)\,. \]
Solutions to (\ref{F.1}) can be written as convolutions
\[ \psi_{\alpha}(x,t)\,=r_{\scriptscriptstyle D}\,
\int^{t}_{0} d \tau \int d^{4}x'\,
G_{\alpha \rho}(x-x',t-\tau )
(\xi_{\rho}(x',\tau)+\eta_{\rho}(x',\tau))  \]
\begin{equation}
\label{F.6}
\end{equation}
\[ \psib_{\alpha}(x,t)\,=r_{\scriptscriptstyle D}\,
\int^{t}_{0} d \tau \int d^{4}x'\,
(\xib_{\rho}(x',\tau)+\etab_{\rho}(x',\tau))
\Gb_{\rho \alpha}(x-x',t-\tau )     \]

The nature of fluctuation terms in (\ref{F.1}) implies that the simplest
nonzero correlation function one can calculate (see(\ref{M.21})) is
\begin{eqnarray}
&&  \langle
\psib_{\alpha}(x_{1},t_{1})\psi^{\alpha}(x_{2},t_{2})
\psib_{\beta}(y_{1},t_{3})\psi^{\beta}(y_{2},t_{4}) \rangle\;\; =
\;r^{4}_{\scriptscriptstyle D}
  \!\! \int^{t_{1}}_{0}\!\!\!\!\!\! d\tau_{1}
\!\!\! \int^{t_{2}}_{0}\!\!\!\!\!\! d\tau_{2}
\!\!\! \int^{t_{3}}_{0}\!\!\!\!\!\! d\tau_{3}
\!\!\! \int^{t_{4}}_{0}\!\!\!\!\!\! d\tau_{4}
\!\!\! \int \!\! d^{4}\!{x'}_{\!\! 1}
\!\!\! \int \!\! d^{4}\!{x'}_{\!\! 2}
\!\!\! \int \!\! d^{4}\!{y'}_{\!\! 1}
\!\!\! \int \!\! d^{4}\!{y'}_{\!\! 2}  \nonumber \\
&& \langle
(\xib_{\sigma}({x'}_{\!\! 1},\tau_{1})\! + \!
\etab_{\sigma}({x'}_{\!\! 1},\tau_{1}))
\Gb_{\sigma\alpha}(x_{1}\! -\! {x'}_{\!\! 1},t_{1}\! -\! \tau_{1})
G_{\alpha\rho}(x_{2}\! -\! {x'}_{\!\! 2},t_{2}\! -\! \tau_{2})
(\xi_{\rho}({x'}_{\!\! 2},\tau_{2})\! + \!
\eta_{\rho}({x'}_{\!\! 2},\tau_{2}))
\nonumber \\
&&  (
\xib_{\mu}({y'}_{\!\! 1},\tau_{3})\! + \!
\etab_{\mu}({y'}_{\!\! 1},\tau_{3}))
\Gb_{\mu\beta}(y_{1}\! -\! {y'}_{\!\! 1},t_{3}\! -\! \tau_{3})
G_{\beta\nu}(y_{2}\! -\! {y'}_{\!\! 2},t_{4}\! -\! \tau_{4})
(\xi_{\nu}({y'}_{\!\! 2},\tau_{4})\! + \!
\eta_{\nu}({y'}_{\!\! 2},\tau_{4})) \rangle
\nonumber \\
& \!\! = & r^{4}_{\scriptscriptstyle D}
\!\! \int^{t_{1}}_{0}\!\!\!\!\!\! d\tau_{1}
\!\! \int^{t_{2}}_{0}\!\!\!\!\!\! d\tau_{2}
\!\! \int^{t_{3}}_{0}\!\!\!\!\!\! d\tau_{3}
\!\! \int^{t_{4}}_{0}\!\!\!\!\!\! d\tau_{4}
\!\! \int \!\! d^{4}\!{x'}_{\!\! 1}
\!\! \int \!\! d^{4}\!{x'}_{\!\! 2}
\!\! \int \!\! d^{4}\!{y'}_{\!\! 1}
\!\! \int \!\! d^{4}\!{y'}_{\!\! 2}   \label{F.7}    \\
& & \!\!\!\!\!\!
\Gb_{\sigma\alpha}(x_{1}\! -\! {x'}_{\! 1},t_{1}\! -\! \tau_{1})
G_{\alpha\rho}(x_{2}\! -\! {x'}_{\! 2},t_{2}\! -\! \tau_{2})
\Gb_{\mu\beta}(y_{1}\! -\! {y'}_{\! 1},t_{3}\! -\! \tau_{3})
G_{\beta\nu}(y_{2}\! -\! {y'}_{\! 2},t_{4}\! -\! \tau_{4})
\nonumber \\
&& \langle
(\xib_{\sigma}({x'}_{\!\! 1},\tau_{1})\! + \!
\etab_{\sigma}({x'}_{\!\! 1},\tau_{1}))
(\xi_{\rho}({x'}_{\!\! 2},\tau_{2})\! + \!
\eta_{\rho}({x'}_{\!\! 2},\tau_{2}))
\times \nonumber \\
& & \;\;\;\;\;\;\;\;\;\;\;\;\;\;\;\;\;\;\;\;\;\;\;\;\;\;\;\;
\;\;\;\;\;\;\;\;\;\;\;\;\;\;\;\;\;\;\;\;\;\; \times
(\xib_{\mu}({y'}_{\!\! 1},\tau_{3})\! + \!
\etab_{\mu}({y'}_{\!\! 1},\tau_{3}))
(\xi_{\nu}({y'}_{\!\! 2},\tau_{4})\! + \!
\eta_{\nu}({y'}_{\!\! 2},\tau_{4})) \rangle
\nonumber
\end{eqnarray}
where $\langle...\rangle$ represents integration over Grassmann variables.
Due to the properties of stochastic variables (\ref{M.24},\ref{M.25}),
integral (\ref{F.7}) over region
$t_{1}\,\times\,t_{2}\,\times\,t_{3}\,\times\,t_{4}$ is equal to zero.
To stay on the manifold where the mean value $\langle...\rangle\,\neq\,0$,
it is necessary to impose condition that the time integrations are carried
out only for $t_{1}\,=\,t_{2}\,\leq\,t_{3}\,=\,t_{4}$. Hence (\ref{F.7})
will be first integrated with the weight
\begin{equation}
c_{\tau}\:\delta(\tau_{1}-\tau_{2})\, \delta(\tau_{3}-\tau_{4})\,,
\label{F.8}
\end{equation}
where $c_{\tau}=1,\,[c_{\tau}]=m^{-2}$. The numerical value of $c_{\tau}$
can be again adjusted to $1$ by means of $r_{\scriptscriptstyle D}$ constant
as in all expressions leading to measurable quantities one has always the
product $r^{4}_{\scriptscriptstyle D}\,c_{\tau}\,c_{\scriptscriptstyle E}$
which is dimensionless. The condition (\ref{F.8}) means, that the product
$\psib_{\alpha}\psi^{\alpha}$ will be always taken in the same stochastic
time instant. From (\ref{F.7}) one arrives at
\begin{eqnarray}
\langle\psib_{\alpha}\psi^{\alpha}\psib_{\beta}\psi^{\beta} \rangle& = &
\,2c_{\tau}c_{\scriptscriptstyle E}r^{4}_{\scriptscriptstyle D}
    \int^{\infty}_{0}\!\!\!\!\!d\tau_{1}
\!\!\int^{\infty}_{0}\!\!\!\!\!d\tau_{3}
\!\!\int \!\!d^{4}{x'}_{1} \!\!\int d^{4}{y'}_{1}
\!\!\int\frac{d^{4}p_{1}}{{(2\pi)}^{4}}
\!\!\int\frac{d^{4}p_{2}}{{(2\pi)}^{4}}
\!\!\int\frac{d^{4}p_{3}}{{(2\pi)}^{4}}
\!\!\int\frac{d^{4}p_{4}}{{(2\pi)}^{4}} \nonumber \\
&\Theta&\!\!\!\!\!(t_{1}-\tau_{1})
\exp [-(\psla_{1}+m)(t_{1}-\tau_{1})-ip_{1}x_{1}]_{\sigma\alpha}
\nonumber \\
&\Theta&\!\!\!\!\!(t_{2}-\tau_{1})
\exp [-(\psla_{2}+m)(t_{2}-\tau_{1})+ip_{2}x_{2}]_{\alpha\rho}
\exp [i(p_{1}-p_{2}){x'}_{1}]    \nonumber \\
&\Theta&\!\!\!\!\!(t_{3}-\tau_{3})
\exp [-(\psla_{3}+m)(t_{3}-\tau_{3})-ip_{3}y_{1}]_{\mu\beta}
\nonumber      \\
&\Theta&\!\!\!\!\!(t_{4}-\tau_{3})
\exp [-(\psla_{4}+m)(t_{4}-\tau_{3})+ip_{4}y_{2}]_{\beta\nu}
\exp [i(p_{3}-p_{4}){y'}_{1}]    \nonumber \\
&\delta_{\rho\mu}&\!\!\!\delta_{\sigma\nu}\,
\exp [ \pm M \mid \tau_{3}-\tau_{1} \mid \,]
\label{F.9}
\end{eqnarray}
carrying out also the integrations over $\delta$-functions
\[ \delta({x'}_{1}-{x'}_{2})\, \delta({y'}_{1}-{y'}_{2}) \]
which come from (\ref{M.21}). Further integrations lead to
\begin{eqnarray}
\langle\psib_{\alpha}\psi^{\alpha}\psib_{\beta}\psi^{\beta}\rangle\,
=\,2r\!\!\int^{\infty}_{0}\!\!\!\!\!d\tau_{1}
\!\!\int^{\infty}_{0}\!\!\!\!\!d\tau_{3}
\!\!\int\frac{d^{4}p_{1}}{{(2\pi)}^{4}}
\!\!\int\frac{d^{4}p_{3}}{{(2\pi)}^{4}}\,\,
\Theta(T_{1}-\tau_{1})\Theta(T_{3}-\tau_{3})  \nonumber \\
\exp [-2(\psla_{1}+m)(T_{1}-\tau_{1})]_{\sigma\rho}
\exp [-2(\psla_{3}+m)(T_{3}-\tau_{3})]_{\rho\sigma} \nonumber  \\
\exp [i(x_{2}-x_{1}){p}_{1}] \,\exp [i(y_{2}-y_{1}){p}_{3}] \,
\exp [\pm M \mid \tau_{3}-\tau_{1} \mid \,] \,
\label{F.10}
\end{eqnarray}
where $r\;=\;r^{4}_{\scriptscriptstyle D}\,c_{\tau}\,
c_{\scriptscriptstyle E}$ and it is dimensionless. Further
\[ T_{1}\,=\,min(t_{1},t_{2})\,,\,T_{3}\,=\,min(t_{3},t_{4})\,. \]

So far only $\delta$-functions have been integrated. To integrate over
remaining stochastic time variables, one needs to prove convergence of the
integrals (\ref{F.10}). This is easily done realizing that there
exists a unitary transformation which diagonalizes the matrix $(\psla +
m)$ \cite{JAP5}.
\begin{equation}
(\psla + m)\,=\,{\cal U}^{-1}(p) \left( \begin{array}{cccc}
i \sqrt{p^{2}}+m & 0 & 0 & 0 \\
0 & i \sqrt{p^{2}}+m & 0 & 0 \\
0 & 0 & -i \sqrt{p^{2}}+m & 0 \\
0 & 0 & 0 & -i \sqrt{p^{2}}+m   \end{array} \right )
{\cal U}(p)\,.
\label{F.11}
\end{equation}

After integrations over $\tau_{1}$ and $\tau_{3}$, where it is necessary to
integrate separately and then to add contributions from the regions
$\tau_{3}\,>\,\tau_{1}$ and $\tau_{1}\,>\,\tau_{3}$, one arrives at
\begin{eqnarray}
\langle\psib_{\alpha}\psi^{\alpha}\psib_{\beta}\psi^{\beta}\rangle\,
=\,2r\!\!\int\frac{d^{4}p_{1}}{{(2\pi)}^{4}}
\!\!\int\frac{d^{4}p_{3}}{{(2\pi)}^{4}}\,\,
\exp [i(x_{2}-x_{1}){p}_{1}] \,\exp [i(y_{2}-y_{1}){p}_{3}] \,
\: \times \nonumber \\
\times\: {Tr} \Biggl\{
\exp [ -(2\psla_{1}+2\psla_{3}+4m)T]
\hspace*{\fill}   \nonumber \\
\left( \frac{\fex -1}{(2\psla_{1}+2\psla_{3}+4m)(2\psla_{3}+2m
\mp M)} - \frac{\exp [(2\psla_{1}+2m \pm M)T] - 1}{ (2\psla_{3}+2m \mp
M)(2\psla_{1}+2m \pm M)} +  \nonumber \right. \\
+ \left. \left.
\frac{\fex-1}{(2\psla_{1}+2\psla_{3}+4m)(2\psla_{1}+2m \mp M)} -
\frac{\exp [(2\psla_{3}+2m \pm M)T] - 1}
{(2\psla_{3}+2m \pm M)(2\psla_{1}+2m \mp M)}
\right) \right\} \, , \nonumber \\
\label{F.12}
\end{eqnarray}
where it has been put $T\,\equiv\,T_{1}\,=\,T_{3}$ in accordance with
standard stochastic quantization procedure \cite{ParW}. To carry out the
limit $T\,\rightarrow\,\infty$ one has again to diagonalize the exponential
operators by the unitary transformation (\ref{F.11}). The result is
\begin{eqnarray}
\langle\psib_{\alpha}\psi^{\alpha}\psib_{\beta}\psi^{\beta}\rangle\,
=\,2r\!\!\int\frac{d^{4}p_{1}}{{(2\pi)}^{4}}
\!\!\int\frac{d^{4}p_{3}}{{(2\pi)}^{4}}\,\,
\exp [i(x_{2}-x_{1}){p}_{1}] \,\exp [i(y_{2}-y_{1}){p}_{3}]
\nonumber  \\
{Tr} \left\{ \frac{1}{\psla_{1}+\psla_{3}+2m}
\left( \frac{1}{2\psla_{3}+2m \mp M}
+\frac{1}{2\psla_{1}+2m \mp M} \right) \right\} \, ,
\label{F.13}
\end{eqnarray}
where the signs $\mp M$ correspond to $e^{\pm M\tau}$ in (\ref{M.20}).
In the case of correlations with $e^{+ M\tau}$ it is necessary to assume
\begin{equation}
2\,m\,\,>\,\,M
\label{F.14}
\end{equation}
to assure convergence of (\ref{F.12}) when $T\,\rightarrow\,+\infty$.

     The result (\ref{F.13}) is a four-point function and it is desirable
to reduce it to a two-point function. To achieve this, new momenta have
to be introduced
\begin{equation}
p^{*}\,=\,p_{1}+p_{3}\,\, , \,\, p\,=\,p_{1}-p_{3}
\label{F.15}
\end{equation}
which substituted to (\ref{F.13}) transform it to
\begin{eqnarray}
\label{F.16}
\langle\psib_{\alpha}\psi^{\alpha}\psib_{\beta}\psi^{\beta}\rangle\,
=\,2r\!\!\int\frac{d^{4}p^{*}}{{(2\pi)}^{4}}
\!\!\int\frac{d^{4}p}{{(2\pi)}^{4}}  \hspace*{\fill}    \\
\exp{[ip^{*}[(x_{2}-x_{1})+(y_{2}-y_{1})]]}
\exp [ip[(x_{2}-x_{1})-(y_{2}-y_{1})]]   \nonumber \\
{Tr} \left\{ \frac{1}{p^{*}+2m}
\left( \frac{1}{p^{*}-p+2m \mp M}
+\frac{1}{p^{*}+p+2m \mp M} \right) \right\} \,.\nonumber
\end{eqnarray}
The space coordinate dependence splits into two parts. In
one exponent one has the sum of relative coordinates of the fields
$\psib $ and $\psi$ - this corresponds to $p^{*}$. In the other exponent
the "relative coordinate of relative coordinates" corresponds to p.
It is useful to introduce also a new set of Euclidean space variables
\begin{eqnarray}
z & = & [(x_{2}-x_{1}) \, - \, (y_{2}-y_{1})] \nonumber \\
u & = & [(x_{2}-x_{1}) \, + \, (y_{2}-y_{1})]  \\
v & = & [\, y_{1}\, +\, y_{2}\, ] \nonumber \\
w & = & [\, x_{1}\, +\, x_{2}\, ] \nonumber
\label{F.17}
\end{eqnarray}
The expression (\ref{F.16}) depends on $z$ and $u$ only. Variables $v$
and $w$ are chosen to complete the set of linearly independent
variables. Alternative choices are
\[ v' \, = \, [\, x_{2}\, +\, x_{1}\, +\, y_{2}\, +\, y_{1}\, ]
\,\, , \,\, w' \, = \, w  \]
or
\[ v''\, = \, v'\;\; ,\;\; w''\, = \, [\, y_{1}\, +\, y_{2}\, ]\, . \]

     One possible way, how to reduce the number of space variables, is
to integrate (\ref{F.16}) over $u$. This yields
\begin{eqnarray}
\langle\langle
\psib_{\alpha}\psi^{\alpha}\psib_{\beta}\psi^{\beta}
\rangle\rangle\,\equiv \,
\int du\,\langle\psib_{\alpha}\psi^{\alpha}\psib_{\beta}\psi^{\beta}\rangle\,
=\,2r\!\!\int d^{4}p^{*}\!\!\int\frac{d^{4}p}{{(2\pi)}^{4}}\;
\delta(p^{*}) \; e^{ipz}   \nonumber \\
{Tr} \left\{ \frac{1}{p^{*}+2m}
\left( \frac{1}{p^{*}-p+2m \mp M}
+\frac{1}{p^{*}+p+2m \mp M} \right) \right\} \,.
\label{F.18}
\end{eqnarray}
and subsequent integration over $p^{*}$ leads to
\begin{eqnarray}
\langle\langle
\psib_{\alpha}\psi^{\alpha}\psib_{\beta}\psi^{\beta}
\rangle\rangle
\,=\,2r\!\!\int\frac{d^{4}p}{{(2\pi)}^{4}}\;
e^{ipz} \,\left( \frac{2m \mp M}{m} \right)
\frac{1}{p^{2}+(2m \mp M)^{2}} \, .
\label{F.19}
\end{eqnarray}
The further reductions of space variables imposed by conditions $v=w=0$
lead to
\begin{eqnarray}
\langle\langle\psib_{\alpha}(-x)\psi^{\alpha}(x)
\psib_{\beta}(-y)\psi^{\beta}(y)\rangle\rangle
\,=\,N\!\!\int\frac{d^{4}p}{{(2\pi)}^{4}}\; e^{ip(x-y)} \,
\frac{1}{p^{2}+(2m \mp M)^{2}} \, ,
\label{F.20}
\end{eqnarray}
where all coefficients were absorbed into the constant $N$. By means of
the same procedure applied to the conjugated field combinations one gets
\begin{eqnarray}
\langle\langle\psib_{\alpha}(x)\psi^{\alpha}(-x)
\psib_{\beta}(y)\psi^{\beta}(-y)\rangle\rangle
\,=\,N\!\!\int\frac{d^{4}p}{{(2\pi)}^{4}}\; e^{-ip(x-y)} \,
\frac{1}{p^{2}+(2m \mp M)^{2}} \, ,
\label{F.21}
\end{eqnarray}
which is in accordance with the condition that the overall center of
gravity is at rest.

\section{Long time limit}
\label{secL}

     In the previous section an expectation value and its limit when the
stochastic time goes to infinity was calculated. To be able to ascribe
some meaning to this result, one has to know that it is based on proper
long time limit behavior of the whole process. This means - one has to
prove that the distribution of the solutions to the equations (\ref{F.1})
has a limit when the stochastic time goes to infinity and this limit is
unique.

     To achieve this goal the Fokker-Planck equation cannot be used, as
it is derived under assumption that the relaxation time of fluctuations
is very small compared to observation time. The considered stochastic
process of Grassmann numbers allows both signs in the auto-correlation
function $(e^{\pm M\tau})$ and thus it is desirable to keep full control
of the time evolution rather then to use an approximation.
A significant virtue of stochastic quantization is that it offers the
whole arsenal of stochastic differential equations. It is possible to
use the stochastic Liouville equation \cite{Kemp,Kub}. The idea is to use
the well known hydrodynamic equation
$\stackrel{\cdot}{\rho}\,\,=\,\,-\,div\,\rho \stackrel{\rightarrow}{v}$
in the space
of solutions to (\ref{F.1}). The $\rho$ becomes then the density of
solutions $\psi$ and $\psib$ and there exists
a theorem \cite{Kemp,Kub} saying that $\langle\rho (\psi,\psib ,t)\rangle$,
i.e. the solution density averaged over all fluctuations
$\eta,\xi,\etab ,\xib $
is the probability distribution of $\psi,\psib $ at stochastic time $t$.
It is possible to prove the existence and uniqueness of the long time
limit from the properties of such probability distribution.
The Liouville equation in this case reads
\begin{equation}
{ \partial \over \partial t} \,\, \rho (\psi,\psib ,t)
=
 + \frac{\partial}{\partial\,\psi_{\nu}}\,[\stackrel{\cdot}{\psi_{\nu}}
 \rho (\psi,\psib ,t)]
+\frac{\partial}{\partial\,\psib _{\nu}}\,[\stackrel{\cdot}{\psib _{\nu}}
 \rho (\psi,\psib ,t)]        ,
\label{L.1}
\end{equation}
where the plus sign on the right hand side comes from derivation of the
Liouville equation with anticommuting variables \cite{JAP5}.

     Substitution of the right hand sides of (\ref{F.1}) into (\ref{L.1})
yields in p-representation
\begin{eqnarray}
{ \partial \rho \over \partial t } = \left\{ -8m +
(\psla + m)_{\nu\mu} \psi_{\mu} { \partial \over \partial \psi_{\nu} } +
(\qsla + m)_{\nu\mu} \psib_{\nu} { \partial \over \partial \psib_{\mu} }
\right. \nonumber \\
\left.
- r_{\scriptscriptstyle D}\,( {\xi\prime}_{\nu} + {\eta\prime}_{\nu}  )
{ \partial \over \partial \psi_{\nu} }
- r_{\scriptscriptstyle D}\,( {\xib\prime}_{\nu} + {\etab\prime}_{\nu}  )
{ \partial \over \partial \psib_{\nu} }
\right\} \rho (\psi , \psib , t )
\label{L.2}
\end{eqnarray}
where $\xi\prime ,\xib\prime ,\eta\prime$
and $\etab\prime $ are the Fourier pictures of
stochastic terms and they have the same correlations as in (\ref{M.21})
-(\ref{M.26}). The probability distributions (\ref{M.17},\ref{M.18})
have also to be transformed, e.g. the (\ref{M.17}) transforms as follows
\begin{eqnarray}
&{\cal P}&\!\!\!\!(\xi (p,t),\xib (p',t),\eta (p,t),\etab (p',t))  =
{\cal F} [{\cal P}(\xi (x,t),\xib (x',t),\eta (x,t),\etab (x',t))]
\nonumber \\
&=& \int\frac{d^{4}x}{{(2\pi)}^{4}} \!\!
\int\frac{d^{4}x'}{{(2\pi)}^{4}} \;\;e^{ipx} \, e^{-ip'x'} \;\;
{P}(\xi(x,t),\xib (x,t),\eta (x,t),\etab (x,t))\, \delta (x - x')
\nonumber \\
&=&  \frac{1}{{(2\pi)}^{4}} \int\frac{d^{4}x}{{(2\pi)}^{4}} \;\;
e^{i(p - p')x}\;\;{P}(\xi (x,t),\xib (x,t),\eta (x,t),\etab (x,t))
\label{L.3}
\end{eqnarray}
In order to solve (\ref{L.2}) it is necessary to transform it to the
interaction representation. The stochastic time independent part of
the operator on the right hand side of (\ref{L.2}) can be denoted as
\begin{equation}
A_{0} \:=\:  -8m +
(\psla + m)_{\nu\mu} \psi_{\mu} { \partial \over \partial \psi_{\nu} } +
(\qsla + m)_{\nu\mu} \psib_{\nu} { \partial \over \partial \psib_{\mu} }
\label{L.4}
\end{equation}
and the rapidly varying part as
\begin{equation}
A_{1}(t) \:=\:
- r_{\scriptscriptstyle D}\,( {\xi\prime}_{\nu} + {\eta\prime}_{\nu}  )
{ \partial \over \partial \psi_{\nu} }
- r_{\scriptscriptstyle D}\,( {\xib\prime}_{\nu} + {\etab\prime}_{\nu}  )
{ \partial \over \partial \psib_{\nu} }
\label{L.5}
\end{equation}
After transformation
\begin{equation}
\sigma(t) \:=\: e^{-tA_{0}} \rho(t) \;\;,
\;\;\sigma(0) \:=\: \rho(0) \:\def\: a
\label{L.6}
\end{equation}
the equation (\ref{L.2}) turns into
\begin{equation}
\stackrel{\cdot}{\sigma}(t) \:=\: e^{-tA_{0}} A_{1}(t) e^{tA_{0}}
\;\sigma(t) \:=\: -V(t)\;\sigma(t)
\label{L.7}
\end{equation}
where
\begin{equation}
V(t) \:=\: e^{-tA_{0}} \; A_{1}(t) \; e^{tA_{0}} \,.
\label{L.8}
\end{equation}
Solution to (\ref{L.7}) can be written as the time ordered exponential
\begin{equation}
\sigma(t) \:=\: \lceil \exp \{ - \int^{t}_{0} V(t')\;dt' \} \rceil a
\label{L.9}
\end{equation}
which can be expanded as
\begin{eqnarray}
&\!\!\!\sigma(t) = &\!\!\!a - \int^{t}_{0} \!\!\!dt_{1} V(t_{1}) a
+ \int^{t}_{0} \!\!\!dt_{1} \!\!\int^{t_{1}}_{0} \!\!\!dt_{2}
V(t_{1}) V(t_{2}) a
- \int^{t}_{0} \!\!\!dt_{1} \!\!\int^{t_{1}}_{0} \!\!\!dt_{2}
\!\!\int^{t_{2}}_{0} \!\!\!dt_{3} V(t_{1}) V(t_{2}) V(t_{3}) a
\nonumber \\
&&\!\!\!+ \int^{t}_{0} \!\!\!dt_{1} \!\!\int^{t_{1}}_{0} \!\!\!dt_{2}
\!\!\int^{t_{2}}_{0} \!\!\!dt_{3} \!\!\int^{t_{3}}_{0} \!\!\!dt_{4}
V(t_{1}) V(t_{2}) V(t_{3}) V(t_{4}) a - ...
\label{L.10}
\end{eqnarray}

One gets rid of all the terms in (\ref{L.10}) except of those of the order
$4n (n = 0, 1, 2,..)$ by calculation of the average
$\langle \sigma(t) \rangle$ over
Grassmann stochastic variables. The remaining terms are nonzero provided
that they have been averaged in the same way as in (\ref{F.7}) with the
weight (\ref{F.8}). The first two nonzero terms are
\begin{equation}
\langle S_{0} \rangle \:=\: \langle a \rangle \:=\:
\langle \rho(t=0) \rangle \: \equiv \: 1
\label{L.12}
\end{equation}
which is in fact normalization of probability distribution and
\begin{equation}
S_{1} \:=\:   \int^{t}_{0} dt_{1} \int^{t}_{0} dt_{2}
\int^{t}_{0} dt_{3} \int^{t}_{0} dt_{4}
V(t_{1}) V(t_{2}) V(t_{3}) V(t_{4}) a .
\label{L.13}
\end{equation}
256 terms of the product in $S_{1}$ are reduced by integration over
stochastic terms to the expression
\begin{eqnarray}
\!\!\! \langle S_{1} \rangle = 4r {\displaystyle \int^{t}_{0}}
\!\!\!dt_{1}
{\displaystyle
 e^{-t_{1}A_{0}} { \partial \over \partial \psi_{\nu}(p_{1}) }
{ \partial \over \partial \psib_{\mu}(p_{1}) } e^{t_{1}A_{0}}  }
{\displaystyle \!\!\int^{t}_{0}} \!\!\!dt_{3}
{\displaystyle
 e^{-t_{3}A_{0}} { \partial \over \partial \psi_{\mu}(q_{3}) }
{ \partial \over \partial \psib_{\nu}(q_{3}) } e^{t_{3}A_{0}}
e^{\pm M \mid t_{1}-t_{3} \mid}                                 }
\nonumber \\
\!\!\!\!\!\!+\, 4r {\displaystyle \int^{t}_{0}}
\!\!\!dt_{1}
{\displaystyle
 e^{-t_{1}A_{0}} { \partial \over \partial \psib_{\nu}(q_{1}) }
{ \partial \over \partial \psi_{\mu}(q_{1}) } e^{t_{1}A_{0}}  }
{\displaystyle \!\!\int^{t}_{0}} \!\!\!dt_{3}
{\displaystyle
 e^{-t_{3}A_{0}} { \partial \over \partial \psib_{\mu}(p_{3}) }
{ \partial \over \partial \psi_{\nu}(p_{3}) } e^{t_{3}A_{0}}
e^{\pm M \mid t_{1}-t_{3} \mid}                               }
\label{L.14}
\end{eqnarray}
where the time integrations have been carried out with the weight
(\ref{F.8}) and $r \,=\,r^{4}_{\scriptscriptstyle D}\,c_{\tau}\,
c_{\scriptscriptstyle E}$ is dimensionless
as in the (\ref{F.10}). The absolute values of time difference in
(\ref{L.14}) come from the fact that although the time sequence
is ordered, the Markovian process can run in both directions.
To cope further with (\ref{L.14}) it is necessary to commute the
exponentials with the derivatives. This leads to
\begin{eqnarray}
\!\!\! \langle S_{1} \rangle =&
4r {\displaystyle \int^{t}_{0}} \!\!\!dt_{1}
{\displaystyle \int^{t_{1}}_{0}} \!\!\!dt_{3}
&\!\!\!
{ \partial \over \partial \psi_{\nu} }
\exp [t_{1}(\psla_{1}+m \pm {M \over 2})]_{\nu\tau}
\exp [t_{1}(\psla_{1}+m \pm {M \over 2})]_{\omega\mu}
{ \partial \over \partial \psib_{\mu} }
\nonumber \\
&&\!\!\!
{ \partial \over \partial \psi_{\rho} }
\exp [t_{3}(\qsla_{3}+m \mp {M \over 2})]_{\rho\omega}
\exp [t_{3}(\qsla_{3}+m \mp {M \over 2})]_{\tau\sigma}
{ \partial \over \partial \psib_{\sigma} }
\nonumber \\
\!\!\!\!\!\!+&\!\!\!4r {\displaystyle \int^{t}_{0}}\!\!\!dt_{3}
{\displaystyle \int^{t_{3}}_{0}} \!\!\!dt_{1}
&\!\!\!
{ \partial \over \partial \psi_{\nu} }
\exp [t_{1}(\psla_{1}+m \mp {M \over 2})]_{\nu\tau}
\exp [t_{1}(\psla_{1}+m \mp {M \over 2})]_{\omega\mu}
{ \partial \over \partial \psib_{\mu} }
\nonumber \\
&&\!\!\!
{ \partial \over \partial \psi_{\rho} }
\exp [t_{3}(\qsla_{3}+m \pm {M \over 2})]_{\rho\omega}
\exp [t_{3}(\qsla_{3}+m \pm {M \over 2})]_{\tau\sigma}
{ \partial \over \partial \psib_{\sigma} }
\nonumber \\
\!\!\!\!\!\!&\!\!\! + \:\:cc\:(p \leftrightarrow q)&
\label{L.15}
\end{eqnarray}
where one has to integrate separately and add contributions from the two
possible time orderings $t_{1} > t_{3}$ and $t_{3} > t_{1}$. The $cc$
part originates from the second term in (\ref{L.14}).

It is now necessary to calculate an integral of the type
\begin{equation}
{\cal I}^{\beta\delta}_{\alpha\gamma} \:=\:
\int^{t}_{0} d\tau (e^{A\tau})_{\alpha\beta}\:
(e^{B\tau})_{\gamma\delta}.
\label{L.16}
\end{equation}
By integration per partes one gets
\begin{eqnarray}
\int^{t}_{0} d\tau (e^{A\tau})_{\alpha\beta}\,
(e^{B\tau})_{\gamma\delta}
\:=\: (e^{A\tau})_{\alpha\mu}\,A^{-1}_{\mu\beta}\,
(e^{B\tau})_{\gamma\delta}
{ \displaystyle \mid^{t}_{0} } \;-\;
\int^{t}_{0} d\tau (e^{A\tau})_{\alpha\mu}\,A^{-1}_{\mu\beta}
\,(e^{B\tau})_{\gamma\nu}\,B_{\nu\delta}
\end{eqnarray}
and provided that $[A,B]\,=\,0$
\begin{eqnarray}
\int^{t}_{0} d\tau (e^{A\tau})_{\alpha\mu}\,(e^{B\tau})_{\gamma\nu}\;
[\delta_{\mu\beta}\,\delta_{\nu\delta}\;+\;
A^{-1}_{\mu\beta}\,B_{\nu\delta}]
\:=\: (e^{A\tau})_{\alpha\mu}\,A^{-1}_{\mu\beta}\,
(e^{B\tau})_{\gamma\delta}
\mid^{t}_{0}
\end{eqnarray}
which results in
\begin{eqnarray}
\int^{t}_{0} d\tau (e^{A\tau})_{\alpha\beta}\,
(e^{B\tau})_{\gamma\delta}\:=\:
[\,A_{\alpha\mu}\,\delta_{\gamma\nu}\;+\;
\delta_{\alpha\mu}\,B_{\gamma\nu}\,]^{-1}\;
(e^{A\tau})_{\mu\beta}\,(e^{B\tau})_{\nu\delta}\:\mid^{t}_{0}
\label{L.17}
\end{eqnarray}

In order to calculate (\ref{L.15}) one has to substitute
$A\,=\,B\,=\,\psla_{1}+m \pm {M \over 2}$
(resp. $A\,=\,B\,=\,\qsla_{3} + m \mp {M \over 2}$).
At this moment it is useful put $\psla_{1}\,=\,(\psla^{*}+\psla)/2$ and
$\qsla_{3}\,=\,(\psla^{*}-\psla)/2$
and to apply the procedure
outlined in (\ref{F.15} - \ref{F.19}). This is equivalent to substitution
$\psla_{1}\,=\,-\,\qsla_{3}\,\equiv\,\psla /2$.
Application of (\ref{L.17}) to (\ref{L.15}) then leads to
\begin{eqnarray}
\!\!\!\langle S_{1} \rangle \:= \!\!\!\!\!\!&
\:\:\:2r {\displaystyle \int^{t}_{0}} \!\!\!dt_{1}
\!\!\!\!&
{ \partial \over \partial \psi_{\nu} }
\exp [t_{1}({\psla \over 2} + m \pm {M \over 2})]_{\nu\tau}
\exp [t_{1}({\psla \over 2} + m \pm {M \over 2})]_{\omega\mu}
{ \partial \over \partial \psib_{\mu} }
\nonumber \\
&&
{ \partial \over \partial \psi_{\rho} }
\left\{
\exp [t_{1}(-{\psla \over 2} + m \mp {M \over 2})]_{\rho\gamma}
\exp [t_{1}(-{\psla \over 2} + m \mp {M \over 2})]_{\tau\alpha}
\right.
\nonumber \\
&&
\left.
[-\psla_{\gamma\omega}\delta_{\alpha\sigma} \;-\;
\delta_{\gamma\omega}\psla_{\alpha\sigma} \;+\;2 (2m \mp M) \,
\delta_{\gamma\omega} \delta_{\alpha\sigma}]^{-1}
\:+\: {\cal O}^{\sigma\omega}_{\tau\rho}(\exp(0))
\right\} { \partial \over \partial \psib_{\sigma} }
\nonumber \\
\!\!\!\!\!\!+\!\!\!\!\!\!&
\:2r {\displaystyle \int^{t}_{0}}\!\!\!dt_{3}
\!\!\!\!\!\!&
{ \partial \over \partial \psi_{\nu} }
\left\{[\psla_{\nu\tau}\delta_{\gamma\omega} \;+\;
\delta_{\nu\tau}\psla_{\gamma\omega} \;+\;2 (2m \mp M) \,
\delta_{\nu\tau} \delta_{\gamma\omega}]^{-1} \right.
\nonumber \\
&&
\left. \exp [t_{3}({\psla \over 2} + m \mp {M \over 2})]_{\tau\alpha}
       \exp [t_{3}({\psla \over 2} + m \mp {M \over 2})]_{\omega\mu}
\:+\: {\cal O}^{\alpha\mu}_{\nu\gamma}(\exp(0))
\right\} { \partial \over \partial \psib_{\mu} }
\nonumber \\
&&
{ \partial \over \partial \psi_{\rho} }
\exp [t_{3}(-{\psla \over 2} + m \pm {M \over 2})]_{\rho\gamma}
\exp [t_{3}(-{\psla \over 2} + m \pm {M \over 2})]_{\alpha\sigma}
{ \partial \over \partial \psib_{\sigma} }
\nonumber \\
\!\!\!\!\!\!+\!\!\!\!\!&&cc \, (p \leftrightarrow q)
\label{L.18}
\end{eqnarray}
The terms ${\cal O}(\exp(0))$ come from the lower limits of the
integrals.
It is only their behavior as functions of the stochastic time which is
relevant for further calculations. Multiplication of matrices
and subsequent time integration lead to
\begin{eqnarray}
\!\!\!\langle S_{1} \rangle\;\; =&\!\!\!\!
r\,e^{4mt}\; \displaystyle{
{ \partial \over \partial \psi_{\nu} }
{ \partial \over \partial \psib_{\mu} }
\frac{ [ \psla_{\nu\sigma}\psla_{\rho\mu} +
[ p^{2} + 2(m \mp {M \over 2})^{2}]
\delta_{\nu\sigma}\delta_{\rho\mu} ] }
{m\,(2m \mp M)\;[ p^{2} + (m \mp {M \over 2})^{2}] }
{ \partial \over \partial \psi_{\rho} }
{ \partial \over \partial \psib_{\sigma} } }
\nonumber \\
+&\!\!\!\!
{\cal O}^{\mu\sigma}_{\nu\rho}(\exp[t(2\psla+2m\pm M)])
\displaystyle{
{ \partial \over \partial \psi_{\nu} }
{ \partial \over \partial \psib_{\mu} }
{ \partial \over \partial \psi_{\rho} }
{ \partial \over \partial \psib_{\sigma} } }
\nonumber \\
+&\!\!\!\!
{\cal O}^{\mu\sigma}_{\nu\rho}(\exp[t(-2\psla+2m\pm M)])
\displaystyle{
{ \partial \over \partial \psi_{\nu} }
{ \partial \over \partial \psib_{\mu} }
{ \partial \over \partial \psi_{\rho} }
{ \partial \over \partial \psib_{\sigma} } }.
\label{L.19}
\end{eqnarray}
This expression can be written in the form of the integral
\begin{equation}
\langle S_{1} \rangle \;\;= 4m\;\int^{t}_{0} d\tau
\widetilde{ W}^{\mu\sigma}_{\nu\rho}(\tau)
{ \partial \over \partial \psi_{\nu} }
{ \partial \over \partial \psib_{\mu} }
{ \partial \over \partial \psi_{\rho} }
{ \partial \over \partial \psib_{\sigma} }
\label{L.20}
\end{equation}
where
\begin{eqnarray}
\widetilde{ W}^{\mu\sigma}_{\nu\rho}(\tau) =&\!\!\!\!
\displaystyle{ r\,e^{4m\tau}\;
\frac{ [ \psla_{\nu\sigma}\psla_{\rho\mu} +
[ p^{2} + 2(m \mp {M \over 2})^{2}]
\delta_{\nu\sigma}\delta_{\rho\mu} ] }
{m\,(2m \mp M)\;[ p^{2} + (m \mp {M \over 2})^{2}] }}
\nonumber \\
+&\!\!\!\!
{\cal O}^{\mu\sigma}_{\nu\rho}
([-2\psla+2m\pm M]\exp[\tau(-2\psla+2m\pm M)])
\nonumber \\
+&\!\!\!\!
{\cal O}^{\mu\sigma}_{\nu\rho}
([2\psla+2m\pm M]\exp[\tau(2\psla+2m\pm M)]).
\label{L.21}
\end{eqnarray}
The equation
\begin{equation}
\langle\sigma \rangle = 1 \:+\: 4m\;\int^{t}_{0} d\tau
\widetilde{ W}^{\mu\sigma}_{\nu\rho}(\tau)
{ \partial \over \partial \psi_{\nu} }
{ \partial \over \partial \psib_{\mu} }
{ \partial \over \partial \psi_{\rho} }
{ \partial \over \partial \psib_{\sigma} }
\label{L.22}
\end{equation}
can be then regarded as truncated expansion of the solution to the
equation
\begin{equation}
\langle\stackrel{\cdot}{\sigma} \rangle\: = \:4m\;
\widetilde{ W}^{\mu\sigma}_{\nu\rho}( t )
{ \partial \over \partial \psi_{\nu} }
{ \partial \over \partial \psib_{\mu} }
{ \partial \over \partial \psi_{\rho} }
{ \partial \over \partial \psib_{\sigma} }\langle\sigma\rangle.
\label{L.23}
\end{equation}
After transformation back from the interaction representation,
the equation for $\langle\rho\rangle$ becomes
\begin{equation}
\langle\stackrel{\cdot}{\rho} \rangle\: =
\:A_{0} \langle \rho(t) \rangle \:+\:4m\;
e^{tA_{0}}\widetilde{ W}^{\mu\sigma}_{\nu\rho}( t )
{ \partial \over \partial \psi_{\nu} }
{ \partial \over \partial \psib_{\mu} }
{ \partial \over \partial \psi_{\rho} }
{ \partial \over \partial \psib_{\sigma} }\;e^{-tA_{0}}\;
\langle \rho(t) \rangle
\label{L.24}
\end{equation}
and after commutation of the field derivatives with the operator
$e^{-tA_{0}}$ one obtains
\begin{equation}
\langle\stackrel{\cdot}{\rho} \rangle\: = \:A_{0}\langle\rho(t)\rangle \:+\:
4m\;W^{\mu\sigma}_{\nu\rho}( t )
{ \partial \over \partial \psi_{\nu} }
{ \partial \over \partial \psib_{\mu} }
{ \partial \over \partial \psi_{\rho} }
{ \partial \over \partial \psib_{\sigma} }\;\langle\rho(t)\rangle
\label{L.25}
\end{equation}
where
\begin{equation}
W^{\mu\sigma}_{\nu\rho}(t)\: =\:
\displaystyle{r\,
\frac{[ \psla_{\nu\sigma}\psla_{\rho\mu} +
[ p^{2} + 2(m \mp {M \over 2})^{2}]
\delta_{\nu\sigma}\delta_{\rho\mu} ] }
{m\,(2m \mp M)\;[ p^{2} + (m \mp {M \over 2})^{2}] }  }
\:+\:
{\cal O}^{\mu\sigma}_{\nu\rho}(\exp[t(-2m\pm M)]).
\label{L.26}
\end{equation}

     To investigate the asymptotic properties of solutions to
(\ref{L.25}) it is necessary to find its eigenstates and corresponding
eigenvalues. To this purpose it is useful to introduce
\[  {\mathbf a}^{\dagger}_{\nu} \:=\: {\cal U}_{\nu\alpha}\,\,\psi_{\alpha}
\;\;\;\;\;\;\;,\;\;\;\;\;\;\;
{\mathbf a}^{\rho} \:=\: { \partial \over \partial \psi_{\beta} }\,\,
{\cal U}^{-1}_{\beta\rho}  \]
\begin{equation}
\label{L.27}
\end{equation}
\[  {\mathbf b}^{\dagger \mu} \:=\: \psib_{\gamma}\,\,
{\cal U}^{-1}_{\gamma\mu}
\;\;\;\;\;\;\;,\;\;\;\;\;\;\;
{\mathbf b}_{\sigma} \:=\: {\cal U}_{\sigma\delta}\,\,
{ \partial \over \partial \psib_{\delta} }  \]
where ${\cal U}$ is the unitary transformation which diagonalizes the
matrix $\psla$ and hence also $W$. The operators (\ref{L.27}) are
anticommuting
\begin{equation}
\{ {\mathbf a}^{\rho}\,,\,{\mathbf a}^{\dagger}_{\nu} \} \:=\:
\delta^{\rho}_{\nu}
\;\;\;\;\;\;\;,\;\;\;\;\;\;\;
\{ {\mathbf b}_{\sigma}\,,\,{\mathbf b}^{\dagger \mu}\ \} \:=\:
\delta_{\sigma}^{\mu} .
\label{L.28}
\end{equation}
To convert (\ref{L.25}) into algebraic equation one has to introduce
further the states
\begin{equation}
{\mathbf a}^{\nu}\,\mid 0 \rangle \:=\: 0
\;\;,\;\;
\langle 0 \mid {\mathbf a}^{\dagger}_{\nu}  \:=\: 0
\;\;,\;\;
{\mathbf b}_{\mu}\,\mid 0 \rangle \:=\: 0
\;\;,\;\;
\langle 0 \mid {\mathbf b}^{\dagger\mu}  \:=\: 0.
\label{L.29}
\end{equation}
The search for the eigenstates of the operator on the right hand
side of (\ref{L.25}) leads to the equation
\begin{equation}
0 \:=\:  -8m +
{\cal M}_{\nu}^{\mu} {\mathbf a}^{\dagger}_{\nu}{\mathbf a}^{\mu} +
{\cal N}_{\nu}^{\mu} {\mathbf b}^{\dagger\nu}{\mathbf b}_{\mu} +
4m\,{\cal W}^{\mu\sigma}_{\nu\rho} {\mathbf a}^{\nu}{\mathbf b}_{\mu}
{\mathbf a}^{\rho}{\mathbf b}_{\sigma} \:\equiv\:{\mathbf R}
\label{L.30}
\end{equation}
where
\[ {\cal M}_{\nu}^{\mu} \:=\:
{\cal U}_{\nu\alpha}(\psla + m)_{\alpha\beta}{\cal U}^{-1}_{\beta\mu}
\:\: , \:\:
{\cal N}_{\nu}^{\mu} \:=\:
{\cal U}_{\nu\alpha}(-\psla + m)_{\alpha\beta}{\cal U}^{-1}_{\beta\mu}\]
and
\[ {\cal W}^{\mu\sigma}_{\nu\rho} \:=\:
{\cal U}_{\nu\alpha} {\cal U}_{\rho\gamma}
\: W^{\beta\delta}_{\alpha\gamma} \:
{\cal U}^{-1}_{\beta\mu}{\cal U}^{-1}_{\delta\sigma} . \]

To diagonalize (\ref{L.30}) it is necessary to introduce the similar
transformation
\begin{equation}
{\mathbf O} \:=\: \exp \left( - \,{\cal W}^{\mu\sigma}_{\nu\rho}
{\mathbf a}^{\nu}{\mathbf b}_{\mu}{\mathbf a}^{\rho}{\mathbf b}_{\sigma}
\right).
\label{L.31}
\end{equation}
This transformation changes the operator ${\mathbf R}$
on the right hand side of (\ref{L.30}) to
\begin{equation}
{\mathbf O}\,\,{\mathbf R}\,\,{\mathbf O}^{-1} \:=\:  -8m +
{\cal M}_{\nu}^{\mu}{\mathbf a}^{\dagger}_{\nu}{\mathbf a}^{\mu} +
{\cal N}_{\nu}^{\mu}{\mathbf b}^{\dagger\nu}{\mathbf b}_{\mu}.
\label{L.32}
\end{equation}
The kets are transformed as
\[ \mid \tilde{P} \rangle \:=\: {\mathbf O}\, \mid P \rangle. \]

The equation (\ref{L.32}) has $2^{8}$ different eigenstates with generally
complex eigenvalues. The real parts of the complex eigenvalues are discreete
equal to $-m,-2m,...-7m$, the imaginary parts are functions of the momentum
$p$. The only two eigenstates with pure real eigenvalues $0$ and $-8m$
are nondegenerate.

The eigenstate corresponding to the eigenvalue
$\lambda_{0} = 0$ can be written as
\begin{equation}
\mid \tilde{P}_{0} \rangle \:=\:
\prod_{\nu =0}^{3}\, {\mathbf a}^{\dagger}_{\nu}
\prod_{\mu =0}^{3}\, {\mathbf b}^{\dagger\mu}\, \mid 0 \rangle.
\label{L.33}
\end{equation}
Thus any solution to (\ref{L.1}) averaged over stochastic terms
in this representation (and this order of approximation)
can be written as linear superposition of eigenstates
\begin{equation}
\mid \langle\tilde\rho (t)\rangle\; \rangle \:=\:
\sum_{i =0}^{255}\, e^{\lambda_{i}t}\,c_{i}\, \mid \tilde{P}_{i} \rangle
\label{L.34}
\end{equation}
and in the long time limit $(t \rightarrow \infty)$ survives only
the first term in the sum (\ref{L.34}). Simultaneously vanishes
the term ${\cal O}^{\mu\sigma}_{\nu\rho}(\exp[t(-2m\pm M)])$
in (\ref{L.26}). To untangle the transformations done, it is necessary
to calculate
\begin{equation}
\mid {P}_{0} \rangle \:=\: {\mathbf O^{-1}}\mid \tilde{P}_{0} \rangle \:=\:
\exp \left( {\cal W}^{\mu\sigma}_{\nu\rho}
{\mathbf a}^{\nu}{\mathbf b}_{\mu}{\mathbf a}^{\rho}{\mathbf b}_{\sigma}
\right)
\prod_{\alpha =0}^{3}\, {\mathbf a}^{\dagger}_{\alpha}
\prod_{\beta =0}^{3}\, {\mathbf b}^{\dagger\beta}\, \mid 0 \rangle.
\label{L.35}
\end{equation}
The result is
\begin{equation}
\mid {P}_{0} \rangle \:=\:f(p)\;
\exp \left[ 2\,\,{({\cal W}^{-1})}^{\mu\sigma}_{\nu\rho}
{\mathbf a}^{\dagger\nu}{\mathbf b}^{\dagger}_{\mu}
{\mathbf a}^{\dagger\rho}{\mathbf b}^{\dagger}_{\sigma}
\right]\, \mid 0 \rangle.
\label{L.36}
\end{equation}
where
\begin{equation}
 f(p)\:=\:\left[
\frac{r^{2}\,[4p^{2} + (2m \mp M)^{2}]^{2}\;+\;
(2m \mp M)^{4}}
{m^{2}\,(2m \mp M)^{2}\;[ p^{2} + (m \mp {M \over 2})^{2}]^{2}} \right]
\label{L.37}
\end{equation}
The transformation back to nondiagonal $\psla$ matrices is
straightforward.

Thus in the long time limit the expectation value of any functional
$\langle{\mathbf F}[\psi,\psib]\rangle$ is given by
\begin{equation}
\langle{\mathbf F}[\psi,\psib]\rangle \:=\ \int {\mathbf F}[\psi,\psib] \;
{P}_{0}[\psi,\psib] \; D\psi D\psib
\label{L.38}
\end{equation}
or
\begin{equation}
\langle{\mathbf F}[\psi,\psib]\rangle \:=\
\frac{ \int {\mathbf F}[\psi,\psib] \;
\exp ( - r^{-1}\,{\mathbf S}[\psi,\psib] )\,D\psi D\psib }
{\int \exp ( - r^{-1}\,{\mathbf S}[\psi,\psib] )\,D\psi D\psib }
\label{L.39}
\end{equation}
where
\begin{equation}
{\mathbf S}[\psi,\psib] \:=\ \psib_{\nu}\psib_{\rho}\,
(\,{1 \over 2} \psla_{\nu\sigma}\psla_{\rho\mu} -
[{1 \over 2} p^{2} + (2m \mp M)^{2}]
\delta_{\nu\sigma}\delta_{\rho\mu} )
\psi_{\mu}\psi_{\sigma}\,.
\label{L.40}
\end{equation}

\section{Conclusions}
\label{secC}

Described method of the stochastic quantization of the fermion field yields
as the expectation value of the combination
$\psib_{\alpha}\psi^{\alpha}\psib_{\beta}\psi^{\beta}$ the
scalar boson propagator. The procedure leading to this result is based on
the concept of Markovian processes of Grassmann variables. The notion of
derivative and integral in the Grassmann number calculus allows not only to
introduce the concept of probability but also to extend it to stochastic
processes. The explicit form of one particular process has been constructed.
This process is analogical to classical Ohrenstein-Uhlenbeck process
in the sense that all higher moments are products of the second moments.
However, contrary to the classical process, in the case of Grassmann
variables both negative and positive exponents in autocorrelation exponential
are admissible. This sign appears in the resulting formulae as the sign
of the term which can be interpreted as the mass defect. The negative
mass defect stems from the process which has the "forbidden" sign.

\vspace{\baselineskip}Calculation
of the field correlations proceeds according to standard
stochastic quantization procedure. The new problem arises with the
necessity to reduce the number of dimensions both of the stochastic time
space and of the Euclidean space. This is connected with transition from a
four point function to a two point function. The reduction of two degrees of
freedom in stochastic time follows quite naturally from the properties of
the probability distribution of the stochastic terms. The manifold where it
is nonzero is twodimensional. Staying on this manifold means, that the
product $\psib_{\alpha}\psi^{\alpha}$ is always taken in the same stochastic
time instant, which is natural if one wants to interpret it as one
particle composed of $\psi$ and $\psib$.

The initial system consisted of four fermions and the final system represents
a boson field. Thus the coordinates of individual constituents disappear
from description of the final system. However, the stochastic quantization
provides no mechanism to do this. Hence the redundant degrees of freedom in
Euclidean space are reduced by an ad hoc procedure which is based on
transformation of the momenta to the center of mass system of
$\psib_{\alpha}\psi^{\alpha}$ and its conjugate (which is identical in this
case). The whole expression
$\langle\psib_{\alpha}\psi^{\alpha}\psib_{\beta}\psi^{\beta}\rangle$
is then integrated over space variable corresponding to
the momentum of the overall center of mass $p^{*}$. Consequently $p^{*}$
appears as an argument of a $\delta$-function and it is set to zero by
integration. The whole system is not considered to be on
mass shell. Two other space variables $v$ and $w$ (\ref{F.17}) are also
set to zero which has the meaning of putting the pair $\psi$ and $\psib$
into one space point $x$ resp. $y$.
As a result the final boson propagator depends only on the distance of
two points in Euclidean space and the systems $\langle..\rangle$ and
$\overline{\langle..\rangle}$
propagate with opposite momenta. The energy-momentum conservation constraint
enters the whole description only in this sense.

\vspace{\baselineskip}The dynamics of the fermion field affected by
stochastic process (\ref{M.17},\ref{M.18}) is described in the long time
limit by (\ref{L.40}). It consists of the part which is up to the coefficient
of the mass term equal to free scalar boson Lagrangian and the term which is
analog to the scalar field term in Stueckelberg's Lagrangian of a massive
vector boson field. The functional (\ref{L.38}) does not act on any arbitrary
scalar function $\phi$ but only on spinor functions. Hence this term will
lead in case of a scalar constructed from spinor functions
(${\mathbf F}[\psib_{\alpha}\psi^{\alpha}\psib_{\beta}\psi^{\beta}]$) to
$Tr(\psla\psla)$ and (\ref{L.40}) will turn to free scalar boson Lagrangian
with no selfinteraction.
In case of other functions constructed from spinors, (\ref{L.39}) gives
generally nonzero results - e.g.
${\mathbf F}[\psib\gamma^{\mu}\psi\,\psib\gamma^{\nu}\psi]$ results in
constant $-\delta^{\mu\nu} / m(2m \mp M)$. However, it has to be stressed
that the probability distribution of stochastic terms is just the most
simple one satisfying conditions of a Markovian process with no ambition
to describe any particular dynamics in the long time limit.

The long time
behaviour of the whole system is studied starting from the expansion of
the solution to the Liuville equation. Thus the quality of approximation
by Fokker-Planck equation can be controlled. The convergence of the
expansion is given by numerical values of $r$ and
fermion mass $m$ as the coefficients corresponding to $m\,(2m \mp M)$
in denominator of $S_{1}$ (\ref{L.19}) become equal to
$m^{2}\,(6m \mp M)(2m \mp M),\;m^{3}\,(10m \mp M)(6m \mp M)(2m \mp M),..$etc
in higher terms $S_{2},\,S_{3},..$. The Gaussian behaviour of stochastic
terms implies that higher order terms should reduce to products of free
"boson Lagrangians" (each containing the term discussed above) with masses
increasing by $4m$. However, then the procedure leading to reduction of the
degrees of freedom has to be generalized and the result will depend on it.

\vspace{\baselineskip}The whole calculation closely follows the standard
stochastic quantization procedure. The only two differences in the approach
consist in different statistical properties of the stochastic terms and in
reduction of the degrees of freedom in final formulae. The later difference
is clearly connected with desired "coarse graining" of the description.
On the other hand the statistical properties of stochastic terms in
Langevin equations are responsible for the fact that in the long (stochastic)
time limit the free fermions disappear and the combination
$\langle\psib_{\alpha}\psi^{\alpha}\psib_{\beta}\psi^{\beta}\rangle$
propagates as a
scalar boson field. Thus the stochastic terms can be regarded as the
external field coupling the fermions together. The question whether this
interaction is an effective result of a standard interaction via gauge fields
lays beyond the scope of the described procedure as the statistical nature
of stochastic terms is imposed ad hoc. Any hint that the given method
reflects real dynamics would generate a puzzling problem - mere statistical
properties of stochastic term make difference between standard stochastic
quantization of free fermion field and description of interacting system
which gets quantized "along the way" in course of calculation.

\vspace{\baselineskip}It is a pleasure to thank Ji\v{r}\'{\i} Rame\v{s} and
Ji\v{r}\'{\i} Ch\'{y}la for reading the manuscript and also to acknowledge
very stimulating discussion with Ji\v{r}\'{\i} Ho\v{r}ej\v{s}\'{\i}.

\end{document}